\documentclass[twocolumn]{aastex63}
\usepackage[mediumspace,mediumqspace,Grey,squaren]{SIunits}

\graphicspath{{./}{figures/}}

\submitjournal{ApJ}

\shorttitle{Stellar populations in NGC 1846}
\shortauthors{Li et al.}

\begin{document}

\title{Multiple stellar populations at less evolved stages-II: no evidence of significant helium spread among NGC 1846 dwarfs}

\correspondingauthor{Chengyuan Li}
\email{lichengy5@mail.sysu.edu.cn}

\author{Chengyuan Li} 
\affil{School of Physics and Astronomy, Sun Yat-sen University, Daxue Road, Zhuhai, 519082, China}
\begin{abstract}
The detection of star-to-star chemical variations in star clusters older than 2 Gyr has changed the traditional view of star clusters as canonical examples of "simple stellar populations" (SSPs) into the so-called ``multiple stellar populations'' (MPs). Although the significance of MPs seems to correlate with cluster total mass, it seems that the presence of MPs is determined by cluster age, however. 
In this article, we use deep photometry from the {\sl Hubble Space Telescope} to investigate whether the FG-type dwarfs in the $\sim$1.7 Gyr-old cluster NGC 1846, have helium spread. By comparing the observation with the synthetic stellar populations, we estimate a helium spread of $\Delta{Y}\sim0.01\pm0.01$ among the main-sequence stars in NGC 1846. The maximum helium spread would not exceed $\Delta{Y}\sim0.02$, depending on the adopted fraction of helium-enriched stars. To mask the color variation caused by such a helium enrichment, a nitrogen enrichment of at least $\Delta{\rm [N/Fe]}$=0.8 dex is required, which is excluded by previous analyses of the red-giant branch in this cluster.
We find that our result is consistent with the $\Delta{Y}$--mass relationship for Galactic globular clusters. To examine whether or not NGC 1846 harbors MPs, higher photometric accuracy is required. We conclude that under the adopted photometric quality, there is no extreme helium variation among NGC 1846 dwarfs. 
\end{abstract}


\section{Introduction} \label{S1}
The phenomenon of chemical variations among stars in star clusters, known as 
multiple stellar populations (MPs), is a very common feature for star clusters \citep{Bast18a}.
Most old globular clusters (GCs) in the Milky Way or the Satellites exhibit MPs \citep[e.g.,][]{Carr09a,Milo20a,Nied17a,Lars14a}, with a small fraction of old GCs have no MPs \citep[e.g.,][]{Vill13a,Dott18a,Milo20a}. Recently, MPs are detected in some intermediate-age clusters (2--6 Gyr-old) in the Magellanic Clouds \citep{Mart18a,Li19a,Holl19a}. But so far clusters younger than $\sim$2 Gyr are not reported to have MPs yet \citep{Mucc08a,deSi09a}. These latter are poorly studied because of the large distance and the few works available focus on the analysis of the evolved giant stars  \citep{Zhang18a,Li19a}. Less-evolved dwarfs in these distant clusters are explored only very recently, based on ultra-deep images \citep{Cabr20a,Li20a,Li21a}. 

Multiple scenarios were proposed to account for the presence of MPs, such as pollutions from massive binaries \citep{deMi09a}, fast-rotating massive stars \citep{Krau13a}, asymptotic giant branch (AGB) stars \citep{Dant16a,Calu19a}, very massive stars \citep[$10^2$--$10^3$ $M_{\odot}$,][]{Vink18a}, or their combination \citep{Wang20a}. The star-to-star chemical variations are usually for light elements, such as He, C, N, O, Na, Mg, Al \citep{Carr09a,Panc17a}\footnote{If considering their iron abundance, most clusters can be treated as mono-metallic \citep{Carr09b}}.Variations of different elements would indicate different nucleosynthesis processes, such as the CNO cycle at $T\sim15-20$ MK, the NeNa chain at $T\sim25-40$ MK, and the MgAl chain at $T\sim40-70$ MK. All these processes, including the 
pp-chain burning, participate in hydrostatic H-burning, leading to stellar He-enrichment. 
If star-to-star chemical variations are produced in the stellar interior, the He spread 
should be the most common signature of MPs, as He is the most direct H-burning product \citep[e.g.,][]{Renz08a,Kim18a}. Therefore, He enrichment should be a typical feature of secondary stellar populations (stars polluted by the pristine population stars), as presumed in numerical simulations \citep[e.g.,][]{Howa19a}. 

The strength of MPs of Galactic globular clusters (GCs) correlates with the cluster 
total mass, both in terms of the variations of the light elements and the He spread \citep{Milo17a,Milo18a}. The same trend also seems to apply to the Large Magellanic Cloud and Small Magellanic Cloud (LMC and SMC) clusters, \citep{Lagi18a,Chan19a,Milo20a},  indicating that cluster total mass plays an important role in the 
strength of MPs. On the other hand, although the classic evolutionary scenario, which 
attributes MPs to the evolutionary mixing process, has been discarded because of the 
detection of MPs among less-evolved stars \citep[e.g.,][]{Carr04a}. The fact that clusters younger 
than $\sim$2 Gyr do not present MPs has led to the speculation that age is a major factor 
determining the presence of MPs.  \citep{Mart18a,Kaps21a}. 

It is proposed that the scenario that MPs is a specific pattern for low-mass stars, possibly relates to some nonstandard evolutionary effects like rotation \citep{Bast18a}. Based on this hypothesis, late-type dwarfs (later than G-type) in star clusters should exhibit star-to-star chemical variations. \cite{Li21a} have shown that the morphology of the MS of NGC 1978, is consistent with MPs with different CNO abundances. However, since NGC 1978 also has MPs among its red-giant stars \citep{Mart17a}, it remains unclear whether the MPs phenomena is an evolutionary feature. One method to examine this hypothesis is to search for MPs among less-evolved dwarfs in clusters without MPs among their giant stars. Recent studies focusing on such a cluster, NGC 419, report no evidence of chemical variations among its MS dwarfs \citep{Cabr20a,Li20a}, at least in terms of the CNO abundances. 

Because of the high ionization energy, direct measurement of He abundance is 
challenging. Only a limited number of stars have been studied in a handful of globular 
clusters. \cite{Mari14a} have studied 21 horizontal branch (HB) stars of the cluster NGC 2808, 
detecting evidence of a He enhancement of $\Delta{Y}=+0.09$. Through analyzing spectra of 
giant stars in $\omega$ Centauri, \cite{Dupr11a,Dupr13a} report a helium difference of $\Delta{Y}\geq0.17$. 
Indirect exploration of stellar He distribution in star clusters requires high-resolution photometry. He-enrichment will lead the radiative opacity to decrease and increases the mean molecular weight. The overall effect is that He-enhanced stars will become hotter and brighter than normal stars at each stage, and they will evolve faster. Specifically, He spread will complicate patterns of different parts in the color-magnitude diagrams (CMDs), including the main-sequence \citep[MS, NGC 2808, ][]{Dant05a,Piot07a}, the horizontal branch \citep[HB,][]{Jang19a}, and the red-giant branch bump \citep[RGBB,][]{Nata11a,Lagi19a}. 

In this work, we study a Large Magellanic Cloud (LMC) cluster near the critical age, NGC 1846 ($\sim$1.7 Gyr-old). We study the morphology of its MS, to examine if He-spread is present among its FG-type dwarfs. In Section \ref{S2}, we introduce the data reduction and the designed method applied in the analysis. Section \ref{S3} contains the main results. We discuss our results in Section \ref{S4}.

\section{Method and Data Reduction} \label{S2}
\begin{figure*}[!ht]
\includegraphics[width=2\columnwidth]{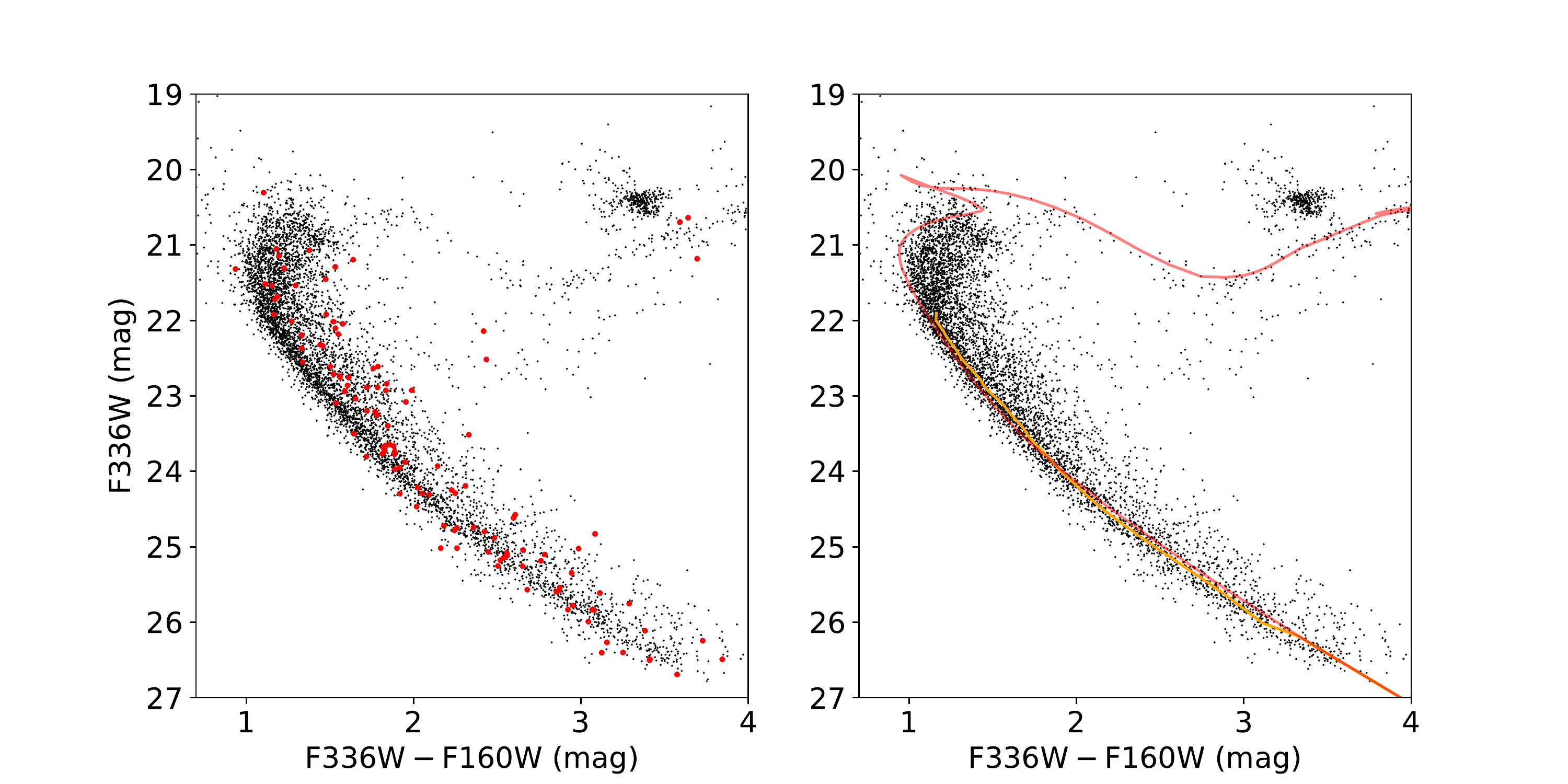}
\caption{The NGC 1846 color-magnitude diagrams (CMDs), red-filled circles are referenced field stars (left panel). Right: the same as the left panel, with the best-fitting isochrone (red solid line) and the calculated main-sequence ridge-line (MSRL, orange solid line).}
\label{F0}
\end{figure*}

Our data is observed through the UVIS-CENTER/WFC3 and IR-FIX/WFC3 from the GO-program 12219 (PI: A.P., Milone), filtered in F336W and F160W passbands with total exposure times of 9156 s and 2844 s, respectively. This wide color baseline allows us to reveal stellar populations with different helium abundances, as He-enriched stars are hotter than normal stars with equivalent masses at the same evolutionary stage \citep{Dant02a}. We perform the point-spread-function (PSF) photometry on the charge transfer efficiency corrected frames (the `{\tt \_flc}' and `{\tt \_drc}' frames), using the {\sc Dolphot2.0} package \citep{Dolp11a,Dolp11b,Dolp13a}. Then we apply standard procedures to the raw stellar catalog to remove bad pixles, saturated objects, extended sources, cosmic rays, and objects hampered by strong crowding \citep[see][]{Li20a}. We present the CMD of NGC 1846 in Fig.\ref{F0}. 

The observed field-of-view (FoV) is limited by the IR-FIX/WFC3, which is 136$\times$123 arcsec$^2$  ($\sim$32.6$\times$29.5 pc$^2$ at the LMC distance). We divide our sample in two parts to minimize the field contamination: the first part represents the region dominated by cluster members \citep[all stars have their distances to the cluster center smaller than $\sim$2 half-light radius, $r\leq{20}$ pc, see][]{Li18a}, another represents the referenced field (beyond $3$ half-light radius, $r\geq{30}$ pc). We remark her that we may have overestimated the field contamination as the size of NGC 1846 is larger than the FoV \citep[its projected tidal radius is $\sim$74 pc, see][]{Li18a}. The CMD of the referenced field is present in Fig.\ref{F0}. 

We calculate the ridge-line of the MS below the turnoff (TO) region (the MS ridge-line, MSRL), using the robust regression method based on the Gaussian process and iterative trimming \citep[ITGP,][]{Li20b,Li20c}\footnote{\url https://github.com/syrte/robustgp/}. Using the MESA Isochrones \& Stellar Tracks (MIST), we generate a series of isochrones with different ages, metallicities and extinctions \citep{Choi16a,Dott16a}. We idenfity the best-fitting isochrones by visually comparing their loci to the calculated MSRL. The best-fitting parameters are $\log({t/{\rm yr}})$=9.24 ($\sim$1.7 Gyr), $Y$=0.26, [Fe/H]=$-$0.64 dex ($Z$=0.008), ($m-M)_0$=18.42 mag (48.31 kpc) and $E(B-V)$=0.024 mag ($A_{V}$=0.08 mag). The best-fitting isochrone and the calculated MSRL are present in the right panel of Fig.\ref{F0}. 

We use the YaPSI isochrone interpolator to evaluate the effect of He variation \citep{Spad17a}\footnote{\url http://www.astro.yale.edu/yapsi/}. Using the best-fitting isochrone as the input of the YaPSI model, we calculate eleven isochrones with He-enrichments of $\Delta{Y}$=0.01--0.11 ($Y$=0.27--0.37). We convert these isochrones into the HST/WFC3 photometric system using the YBC stellar bolometric corrections database \citep{Chen19a}\footnote{\url http://stev.oapd.inaf.it/YBC/}. We calculate their color deviation to the best-fitting isochrone, $\Delta($F336W$-$F160W), and applied the color differences to the calculated MSRL. Finally, we obtain eleven loci for stellar populations with different He-enrichments. 

\begin{figure}[!ht]
\includegraphics[width=\columnwidth]{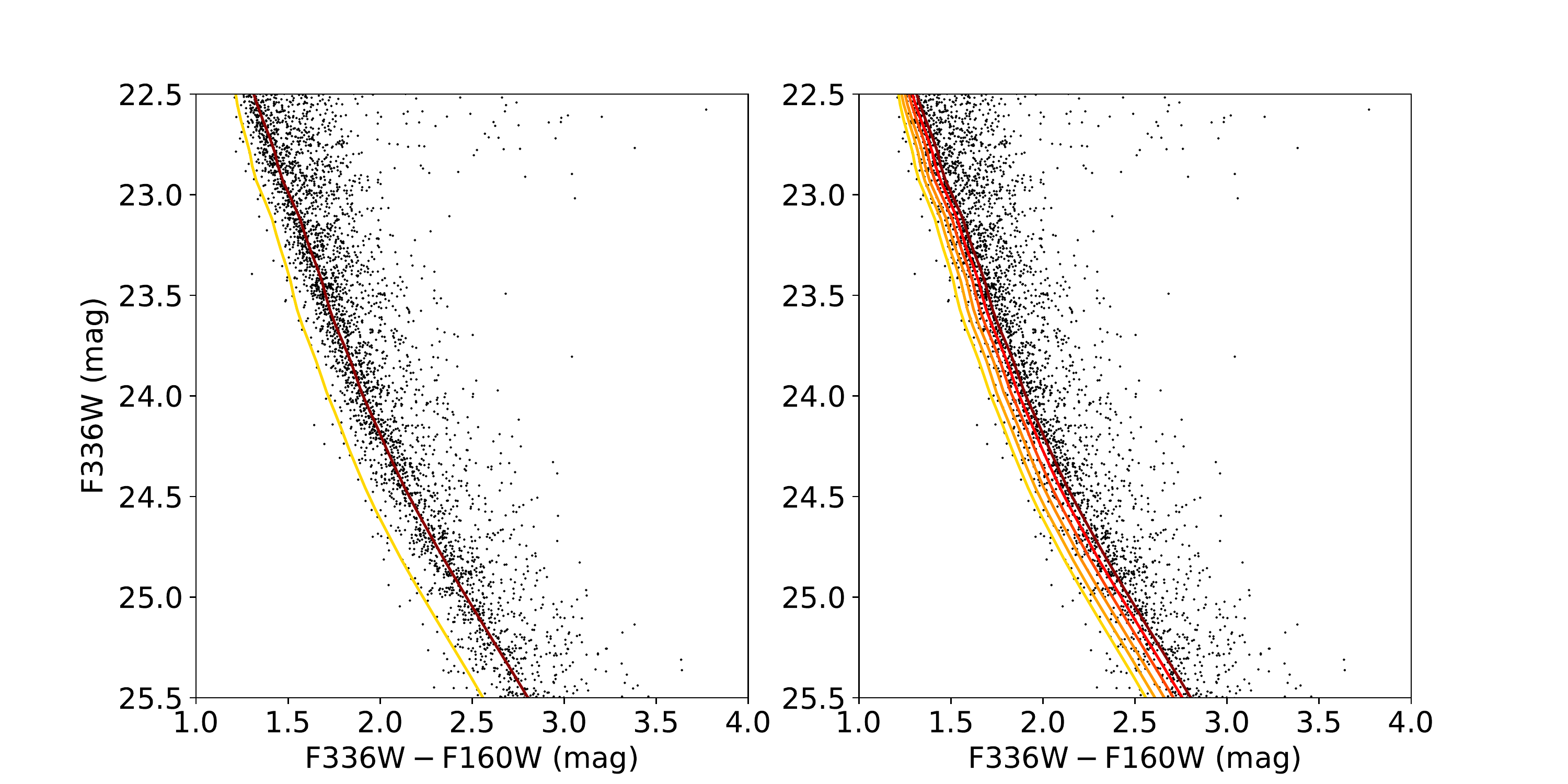}
\caption{Left: color-magnitude diagrams (CMDs) of the NGC 1846, with the best fitting MSRL (brown line) and the He-enriched loci (yellow line, $\Delta{Y}$=0.10). Right: the same as the left panel, but with five He-enriched loci corresponding (from right to left) to $\Delta{Y}$=0.02,0.04,0.06,0.08,0.10.}
\label{F1}
\end{figure}

Fig.\ref{F1} shows the CMD of the MS part with 22.5 mag$\leq$F336W$\leq$25.5 mag, corresponding to a stellar mass range of 1.15$M_{\odot}$ to 0.80$M_{\odot}$ and spectral types from F3--G3. We also present loci of stellar populations with different He abundances. We find that a $\Delta$Y=0.10 would produce a color difference of $\Delta($F336W$-$F160W)$\sim$0.10--0.25 mag, depending on the magnitude range. 

Based on the principle that unresolved MS-MS binaries system will locate in a region that is brighter and redder than the MS, using the same method as in \cite{Li13a}, we derive a MS-MS binary fraction of 20.8\% for NGC 1846 with mass-ratios $q\geq$0.4. Assuming a flat mass-ratio distribution \citep{Li20c}, it indicates a total binary fraction of $\sim$50\%. We assume that this binary fraction does not vary with stellar masses. Finally, we use the MSRL and the different He-abundance loci to generate a series of stellar populations, with 50\% unresolved binaries. At this stage, these generated stellar populations do not contain any photometric uncertainty, differential reddening, and other artifacts such as cosmic rays, hot pixels, and the crowding effect. To mimic the real observation, we produce millions of artificial stars on the MSRL and He-enriched loci. We input these artificial stars into the raw observational frames using the appropriate PSF model, and recover them using the same PSF photometry method applied to real stars. The artificial stellar catalog was further processed follow the same procedures allied to real stars. We randomly select the same number of artificial stars as the real observation, with the same luminosity function. The simulated stellar populations thus contain the photometric uncertainty and artifacts like real stars. However, additional uncertainties may still present due to the differential reddening and zero-point variation \citep{Milo12a}. Following the prescription as in  \cite{Milo12a}, we find that the average differential reddening across the whole FoV is negligible, which are $\Delta{A_{\rm F336W}}$=0.030 mag and $\Delta{A_{\rm F160W}}$=0.004 mag, respectively. The average color variation caused by the zero-point variation and PSF fitting residuals is $\Delta({\rm F336W - F160W})$=0.040 mag. These noises are added to the simulated stellar populations as well. 

\begin{figure}[!ht]
\includegraphics[width=\columnwidth]{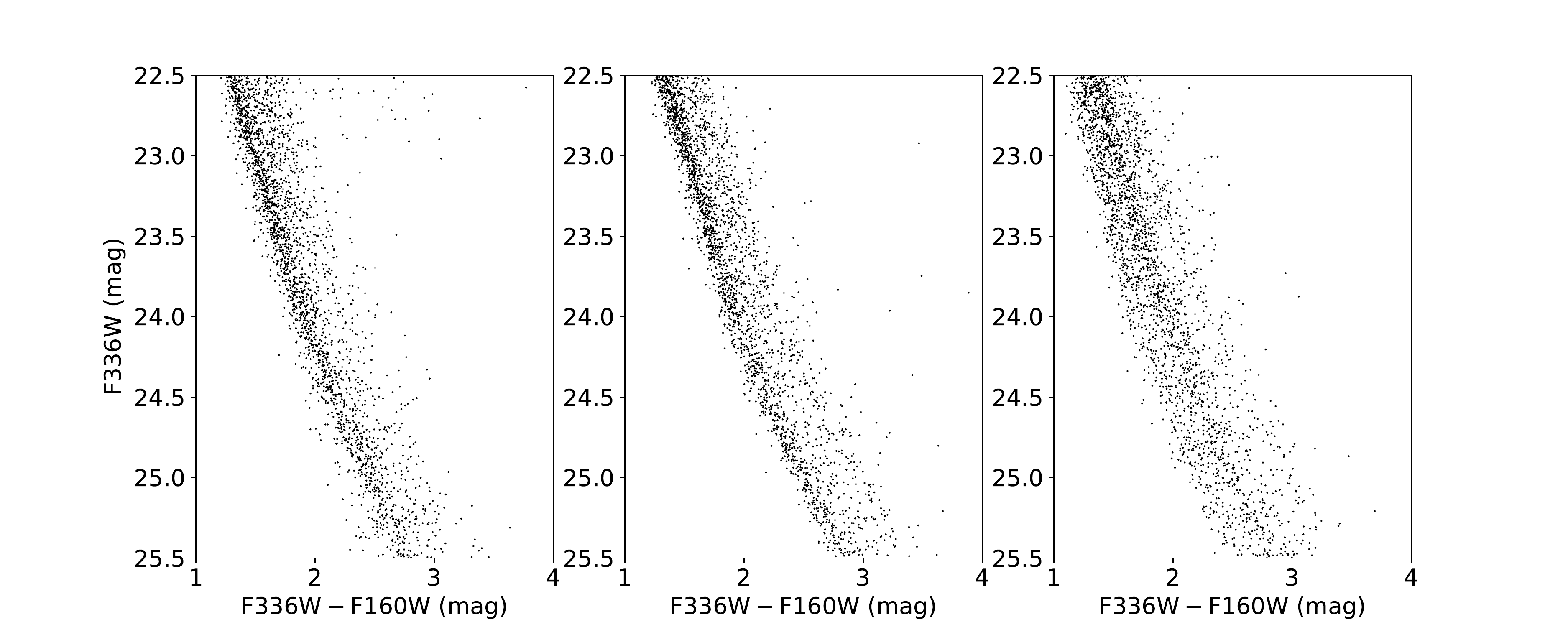}
\caption{Left to right, CMDs of (1) the real observation, (2) the simulated SSP with $Y$=0.26, and (3) the simulated MPs containing $Y$=0.26 and 0.36, respectively.}
\label{F2}
\end{figure}

In Fig. \ref{F2} we show (1) the MS of NGC 1846, (2) the simulated SSP of $Y$=0.26 and (3) the simulated MPs including two populations of $Y$=0.26 and $Y$=0.36 (50\%:50\%), respectively. At first glance, we cannot find a significant difference between observation and the simulated SSP. Nevertheless, a difference appears between the observation and the simulated MPs, as these latter exhibits a clear bifurcation in the MS. 

We first simulate twelve models: a SSP with $Y$=0.26, and eleven MPs with $\Delta{Y}$=0.01--0.11. As a ``toy model'', for MPs, the $Y$ distributions are flat through uniformly mixing stellar populations with different He abundances. As an example, the MPs model with $\Delta{Y}$=0.11 contains twelve populations with $Y$=0.26--0.37, each occupies a number fraction of 1/12 ($\sim$8.3\%). For both the observed and simulated stars, we calculate their color deviations, $\Delta({\rm F336W - F160W})$, to the MSRL. Our goal is to determine the best-fitting model to the observation. We do not adopt any internal CNO variation in our models because of two reasons: (1) \cite{Mart19a} have excluded a significant CNO variation among the red-giant stars of NGC 1846; (2) The analyzed stars are hot enough to destroy the CNO-related molecules (CH and H$_2$O), minimizing the photometric patterns caused by the CNO variation in F336W and F160W passbands. We will discuss the effect of CNO variation in Section \ref{S4}.

As shown in Fig.\ref{F2}, differently to unresolved binaries, He-enriched stars are bluer than the MS. We remove very red stars with $\Delta({\rm F336W - F160W})\geq$0.25 mag to minimize the binary effect. We confirm that $\geq$99\% of these red objects are unresolved binaries. 

We use the two-sample Kolmogorov-Smirnov test (the {\tt KS test2}) to examine the similarity between two stellar populations. {\tt KS test2} is a nonparametric hypothesis test that evaluates the difference between the cumulative distribution functions (CDFs) of the distributions of the two sample data vectors. The input for {\tt KS test2} includes two distributions, the color deviations, $\Delta({\rm F336W - F160W})$, of two population stars. The {\tt KS test2} would return a hypothesis test result--{\tt h}, which is 0 or 1. The returned value of {\tt h} = 1 indicates that {\tt KS test2} rejects the null hypothesis, i.e., the input two samples are not from the same continuous distribution. Otherwise it returns {\tt h} = 0. {\tt KS test2} would also return an asymptotic {\tt p}-value, a scalar value in the range (0,1). The {\tt p}-value is the probability of observing a test statistic as extreme as, or more extreme than, the observed value under the null hypothesis, which is very accurate for large sample sizes for $n_1n_2/(n_1+n_2)\geq4$. In this work, our sample sizes satisfy $n_1n_2/(n_1+n_2)\geq300$. 

We repeat the {\tt KS test2} calculation 50 times for each model to reduce noises caused by the random seeds used in simulated models. The associated {\tt p}-value is usually very low for {\tt h} = 1. As a default rule of thumb, the null hypothesis should be rejected if {\tt p}$<$0.05. To better constrain the difference between models and the observation, we have set a more strict criterion of {\tt p}$<$0.10. We also concern about the {\tt p}-value as a function of $\Delta{Y}$.

\section{Main Results} \label{S3}
\begin{figure*}[!ht]
\includegraphics[width=2\columnwidth]{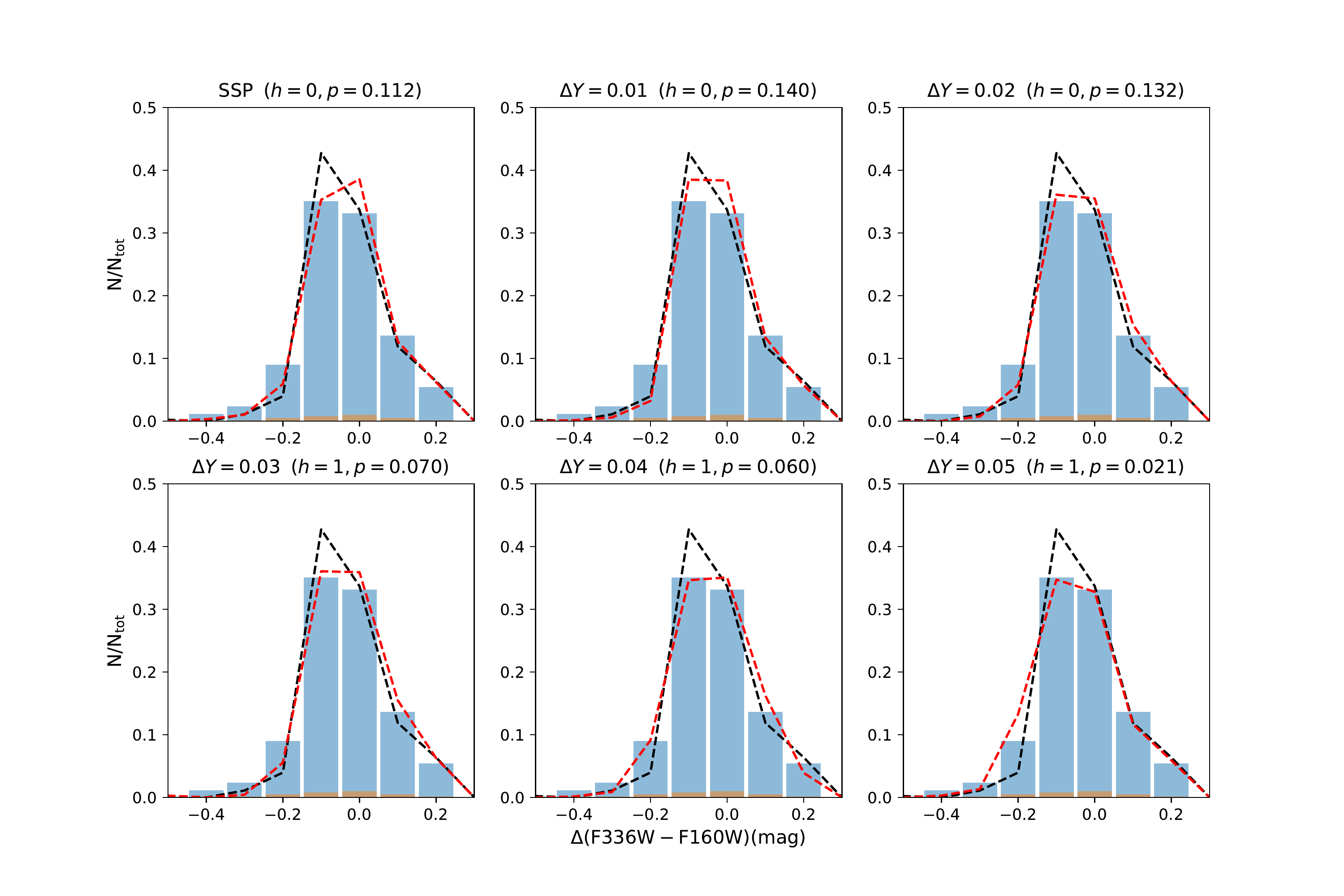}
\caption{The normalized color distribution of the observation (blue and orange histograms are for the cluster and the referenced field), and the simulated population stars (red dashed line). The black dashed line is the expected color distribution for cluster member stars after the field contamination being statistically subtracted. The calculated {\tt h} and {\tt p}-value are are reported in headers.}
\label{F3}
\end{figure*}

\begin{figure*}[!ht]
\includegraphics[width=2\columnwidth]{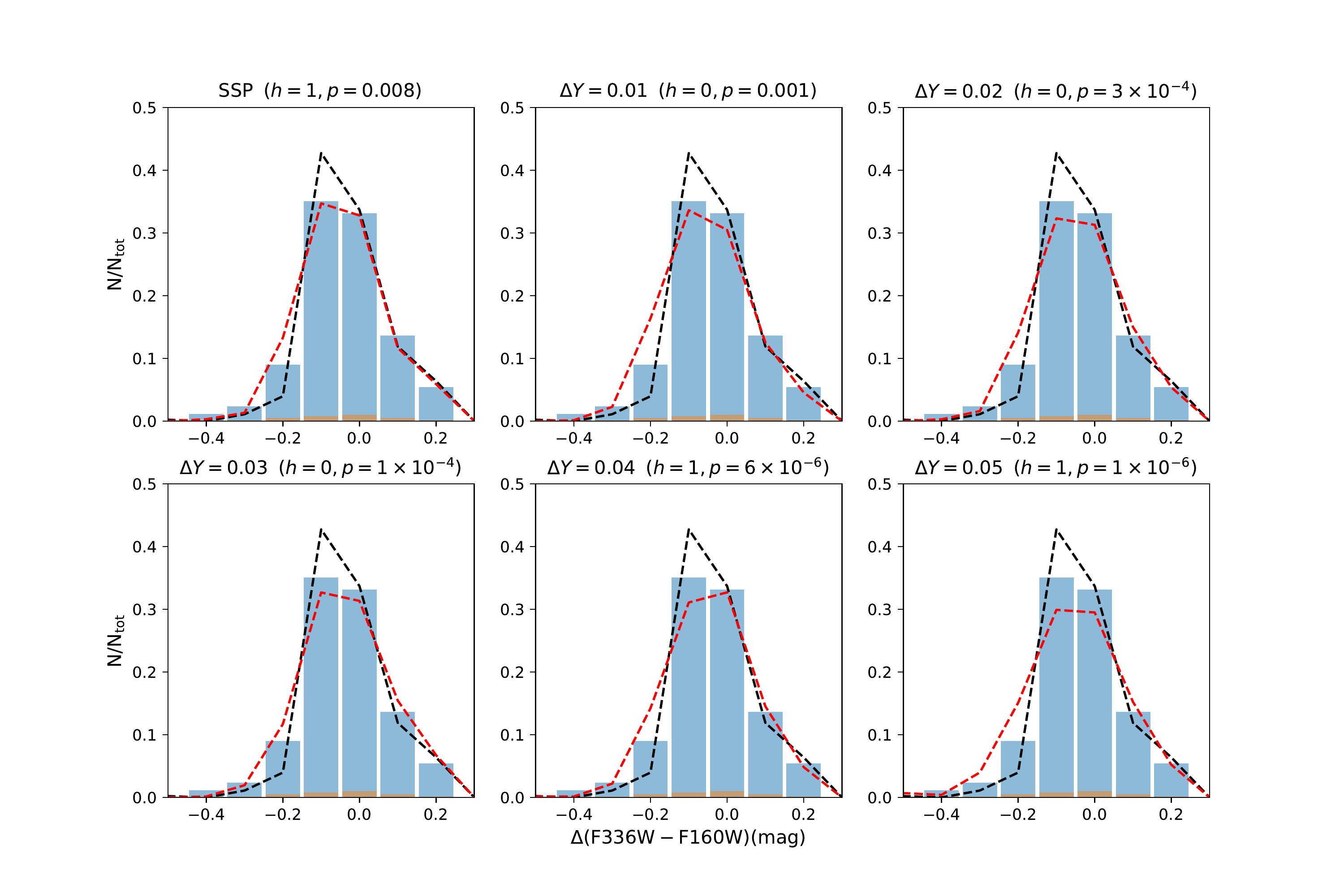}
\caption{The same as Fig.\ref{F3}, but for MPs with $\Delta{Y}$=0.06 to 0.11.}
\label{F4}
\end{figure*}

Fig. \ref{F3} and Fig. \ref{F4} present some pilot results derived from the ``toy model''. They show comparisons of the color distributions between the observation and the simulated stellar populations with different $\Delta{Y}$. The associated {\tt h} and {\tt p}-values calculated by the {\tt KS test2} are also shown in the same figures. We find that the SSP, and MPs with $\Delta{Y}\leq0.03$ can ideally fit the observed color distribution through visually inspect these comparisons. For $\Delta{Y}\geq0.04$, the fitting deteriorates rapidly, however.

Our visual inspection is supported by the {\tt KS test2}. It indicates that the optimal fitting is the model of MPs with $\Delta{Y}$=0.01. Models with $\Delta{Y}$=0.00 (the SSP model) and $\Delta{Y}$=0.02 also returns {\tt p}-values greater than 0.10. For models with $\Delta{Y}\geq0.03$, the {\tt p}-value rapidly decreases. Therefore, the {\tt KS test2} reports that the observed color distribution of the MS is most likely produced by a helium spread of $\Delta{Y}=$0.01$\pm$0.01. 



\begin{figure}[!ht]
\includegraphics[width=\columnwidth]{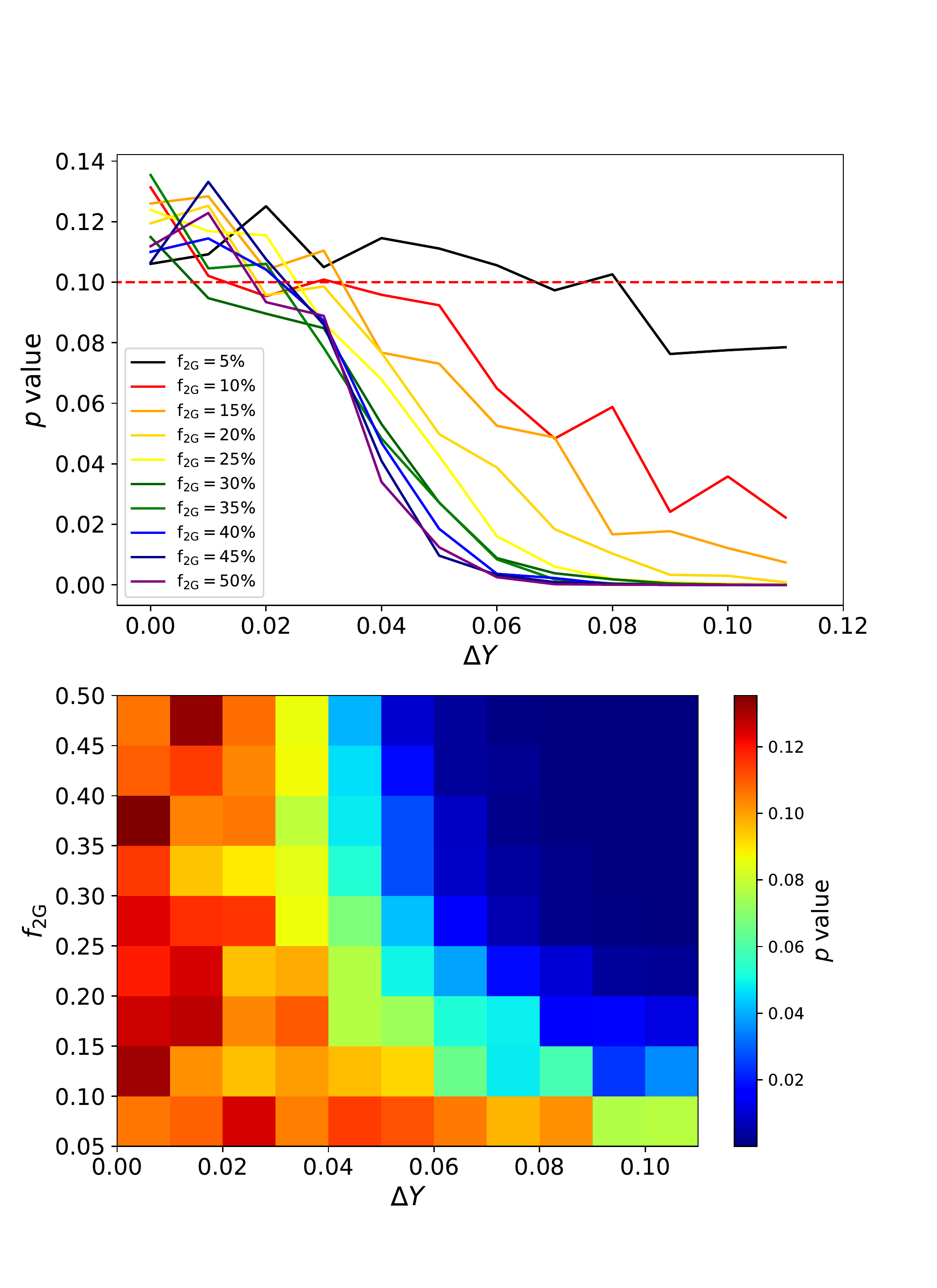}
\caption{Top: The {\tt p}-value as a function of helium spread, $\Delta{Y}$, for models with different 2G fractions. Bottom: 2D distribution of the {\tt p}-value (color-coded), as a function of the 2G fraction ($f_{\rm 2G}$) and helium spread ($\Delta{Y}$). }
\label{F5}
\end{figure}

However, Fig.\ref{F3} and Fig.\ref{F4} only exhibit the results under the extreme cases: in these models, He-enriched population stars always dominated the sample. Several studies have shown that the fraction of chemically enriched stars (2G stars) in the LMC intermediate-age clusters is $\sim$10--20\% \citep[e.g.,][]{Holl19a,Dond21a}. To study the 2G fraction effect, we repeat our simulations for MP models with different fractions of 2G stars, and these results are present in Fig.\ref{F5}. Here we show the {\tt p}-value as a function of $\Delta{Y}$, for models with 2G fractions varying from 5\% to 50\%. Our result shows that except for the model with only 5\% 2G stars, all MP models with $\geq$10\% 2G stars indicate an optimal-fitting helium spread of $\Delta{Y}=0.01\pm0.01$ to the observation. For larger $\Delta{Y}$, the calculated {\tt p}-value rapidly decreases.


\section{Discussion and Conclusion} \label{S4}



\begin{figure}[!ht]
\includegraphics[width=\columnwidth]{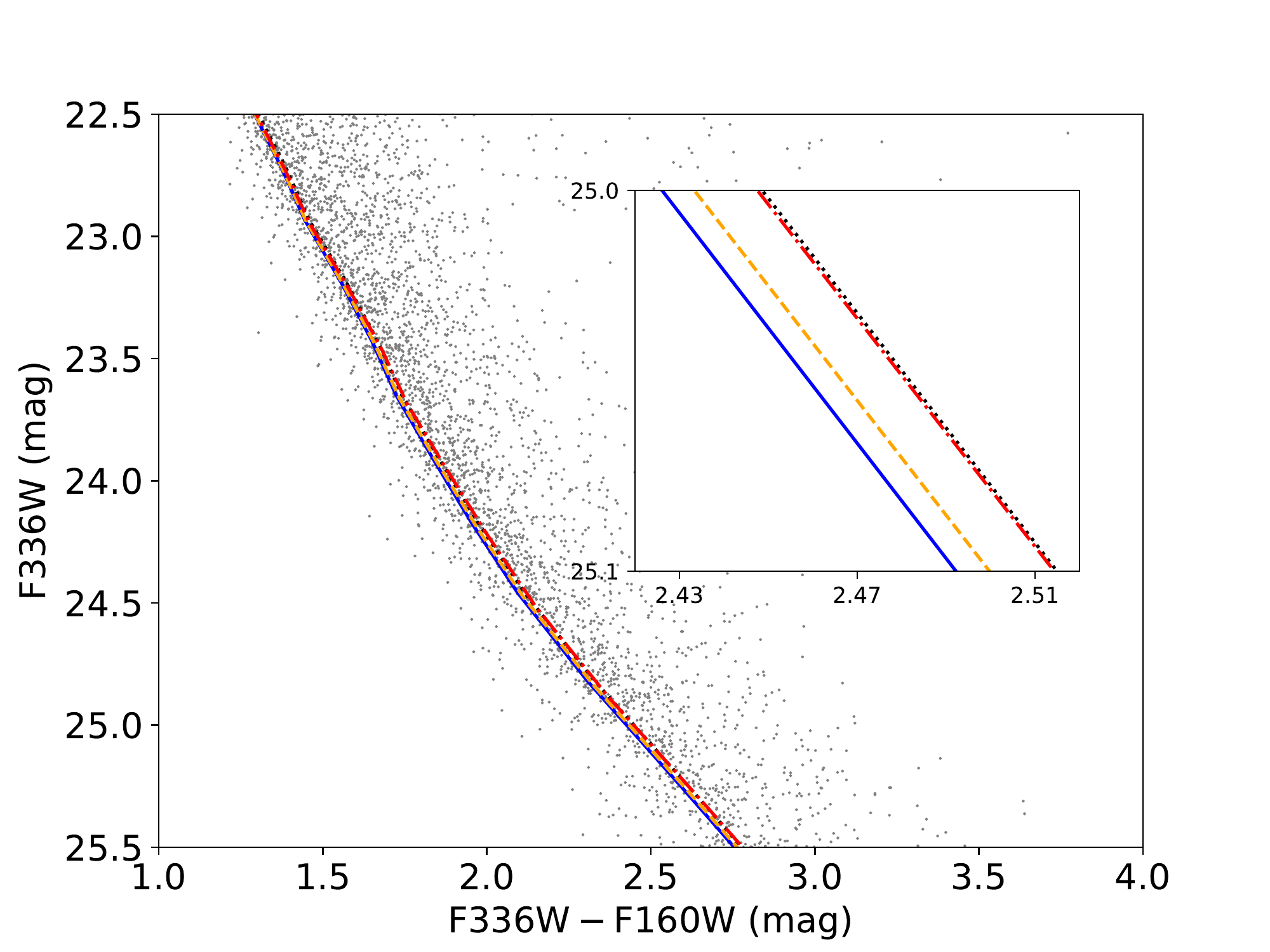}
\caption{The CMD of NGC 1846, as well as sequences of $Y=0.27$ (red dashed line), $Y=0.28$ (blue solid line), $Y=0.28$ with $\Delta{[\rm N/Fe}]$=+0.3 dex (orange dashed line), and $Y=0.28$ with $\Delta{[\rm N/Fe]}$=+0.8 dex (black dotted line).}
\label{F6}
\end{figure}



\subsection{The effect of CNO variation}
As mentioned above, we do not assume any internal CNO variation in our models. In principle, 
stellar photometry in F336W and F160W passbands would be affected by the N and O-related 
molecules. Previous studies have shown that the helium-enriched population stars are enriched in N and depleted in C and O \citep[e.g.,][]{Milo13a,Milo18a}, which 
may reduce the color difference of $\Delta{\rm (F336W-F160W})$ caused by the helium enrichment. 
The enriched N will produce a deeper NH absorption band at 3370\AA, and the 
depleted O will make a shallower H$_2$O absorption at $\sim$1.6 $\mu$m. As a result, the N-enriched (and O-depleted) stellar population will be fainter in F336W, and brighter in F160W than normal stars, making them redder in the color of F336W$-$F160W. Since helium enrichment makes stars hotter (thus bluer), if the helium enriched stars are also N-enriched, they may exhibit negligible color differences to normal stars.

As our result shows a maximum helium variation of $\Delta{Y}=0.01\pm0.01$ ($Y$=0.27$\pm$0.01), we examine how much an increase of the N abundance (and decrease of the C and O abundances) for a Y=0.28 stellar population could produce a similar locus for a Y = 0.27 population. 
To study the effect of CNO variation on the $Y$=0.28 population, we use the method adopted by \cite{Li21a}. Using the global parameters ($\log{\rm g}$, $T_{\rm eff}$, 
$\log{L}$, $X$,$Y$,$Z$) based on the $Y$=0.28 locus, we generate a series of synthetic spectra using 
the package {\sc Spectrum 2.77}\footnote{\url http://www.appstate.edu/~grayro/spectrum/spectrum.html} \citep{Gray94a}, adopting ATLAS9 stellar atmosphere models \citep{Cast03a}. We then generate another series of synthetic spectra with identical parameters but enriched N and depleted C and O (the total abundance of the CNO remains unchanged). For each pair, we convolve their spectra with the F336W and F160W passbands and calculated their flux ratio. Finally, we convert this ratio into magnitude differences in F336W and F160W passbands and used them to correct the $Y$=0.28 locus. We then examine if the N-enriched, $Y$=0.28 stellar locus would overlap with the $Y$=0.27 population (without CNO variation) in the CMD.

According to \cite{Mart19a}, there is an upper limit of N-enrichment of $\Delta{\rm [N/Fe]}<$+0.3 dex for NGC 1846. We therefore first examine the effect of a CNO variation with $\Delta{\rm [N/Fe]}=$+0.3 dex and $\Delta{\rm [C/Fe]}=\Delta{\rm [O/Fe]}$=$-$0.04 dex. The result is present in Fig.\ref{F6}. We find that an N-enrichment of $\Delta{\rm [N/Fe]}=$+0.3 dex will slightly shift the $Y$=0.28 population towards the red direction, but this effect cannot mask the color difference between $Y$=0.28 and $Y$=0.27 populations.

Next, we remove the constraint of N-enrichment to examine at what level the N-enrichment will fully mask the color difference caused by the helium enrichment. We find that if the $Y$=0.28 population is N-enriched with $\Delta{\rm [N/Fe]}=$+0.8 dex ($\Delta{\rm [C/Fe]}=\Delta{\rm [O/Fe]}$=$-$0.26 dex), it becomes indistinguishable from the $Y$=0.27 population. The effect of CNO variation is small for our stars of interest. Because these stars are FG-stars (F3--G3 type) with surface temperatures of at least 5700 K. At these temperatures, the CNO related molecule (NH, H$_2$O) absorption features are too weak to affect the photometry at F336W and F160W passbands.

In summary, our result indicates that NGC 1846 harbors helium spread that does not exceed $\Delta{Y}$=0.01. If it contains MPs, it may also contain an N-enrichment of at least $\Delta{\rm [N/Fe]}=$+0.8 dex, which is excluded by \cite{Mart19a}.

\begin{figure}[!ht]
\includegraphics[width=\columnwidth]{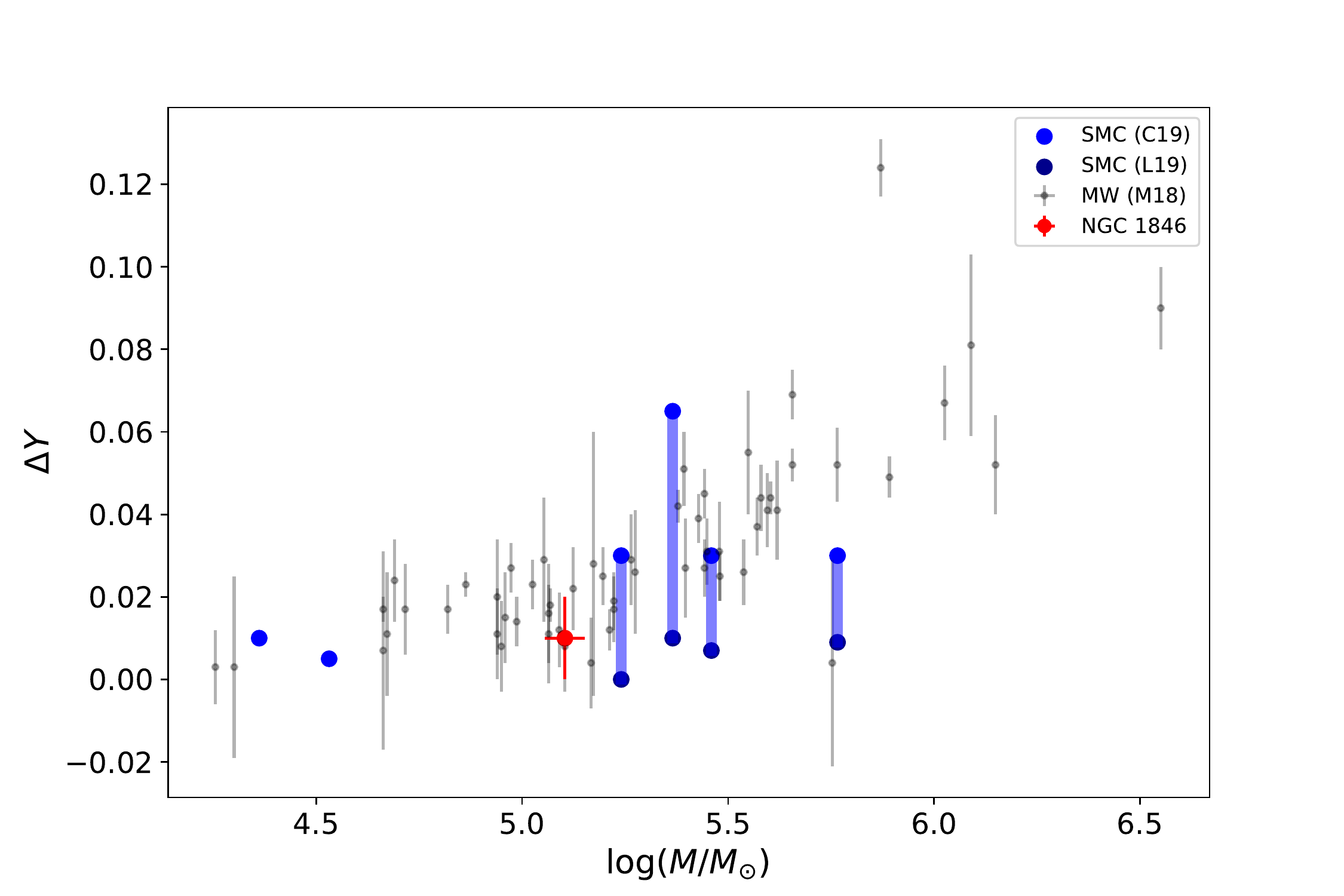}
\caption{Relation between the maximum helium spread, $\Delta{Y}$, and cluster total masses. Blue circles are six SMC clusters from \citep[C19]{Chan19a}, four of them are studied in \citep[dark blue circles,][L19]{Lagi19a} as well. Black dots are Galactic GCs \citep[M18]{Milo18a}. Red circles are NGC 1846 analyzed in this work.}
\label{F7}
\end{figure}

\subsection{The role of the cluster total mass}
A disadvantage of our method is that it is not sensitive to the fraction of 2G stars. As shown in Fig.\ref{F5}, the calculated {\tt p}-value does not vary significantly with the 2G fraction. Because of this, we cannot derive an optimal 2G fraction for observations. According to \cite{Dond21a}, if NGC 1846 has MPs and its 2G fraction is no less than 10\%, we can exclude the possibility that NGC 1846 has an extreme helium spread ($\Delta{Y}\geq0.04$) among its dwarf stars. Its internal helium spread is most likely $\Delta{Y}=0.01$ and does not exceed $\Delta{Y}=0.02$. However, our result cannot rule out the possibility that NGC 1846 is an SSP cluster because of the accuracy of the {\tt KS test2} calculation.

Similar to Galactic GCs \citep{Milo18a}, \cite{Chan19a} find that the maximum helium spreads of Magellanic Clouds clusters are correlated with the total cluster masses, although in their work they did not contain any intermediate-age cluster. However, their helium determination from the HB morphology may only represent an upper limit, because the fact that 2G stars may have experienced more mass loss than the 1G is not taken into considered \citep[e.g.,][]{Tail20a}. The helium spreads of SMC clusters derived from their RGB morphologies are smaller, indeed \citep[see Fig.\ref{F7}]{Lagi19a}. Our helium determination is thus more realistic as it is derived from the cluster's MS morphology. To compare our result with the derived correlation for old GCs, we use the expected mass at $\sim$10 Gyr for NGC 1846 derived by \cite{Goud11a}. We confirm that  a helium spread of $\Delta{Y}=0.01\pm0.01$ for an NGC 1846-like GC is entirely consistent with the correlation between the maximum helium spread and clusters total masses (Fig.\ref{F7}). If the formation of MPs is a standard process of massive star clusters, the absence of a significant helium 
spread in NGC 1846 is because it is not sufficiently massive.

\subsection{The role of the cluster age}
\cite{Milo20a} suggest an initial mass threshold of $\sim$1.5$\times10^5$ $M_{\odot}$ for the presence of MPs. The initial mass of NGC 1846 may range from 2$\times10^5$ $M_{\odot}$ to 6$\times10^5$ $M_{\odot}$, depending on its initial mass segregation degree \citep{Goud14a}. If NGC 1846 is a SSP cluster, its lack of MPs may support the possibility that age controls the appearance of MPs. However, the initial mass estimation of NGC 1846 is still uncertain, the difference between its estimated initial mass and the suggested mass threshold can be explained by the systematic errors \citep{Milo20a}.

We can compare the current mass of NGC 1846 to MC clusters known to have MPs with similar ages. \cite{Chan19a} have studied the helium spread of Hodge 6 and NGC 1978, other two intermediate-age LMC clusters, through analyzing their HB morphologies. The mass of NGC 1846 is $\sim$60\% of NGC 1978 and about twice the Hodge 6, but they are similar in age (NGC 1846 is about 200--300 Myr younger). \cite{Chan19a} cannot conclude whether these two clusters harbor helium spread because of the uncertainty of the mass loss role for these two clusters. They suggest that the maximum helium spreads for Hodge 6 and NGC 1978 would be $\Delta{Y}\sim$0.06 and  $\Delta{Y}\sim$0.04, respectively, once they assume a constant mass loss coefficient, $\eta_{\rm R}$. If their assumption holds, and if NGC 1846 is an SSP cluster, it may imply the importance of age for the formation of MPs. The critical age range in which clusters begin to transit from SSPs to MPs is $\sim$1.7--2.0 Gyr. If age does plays a role on the properties of MPs, it should only important for the onset of MPs. As \cite{Lagi19a} have excluded a strong correlation between the strength of MPs and the age of the host cluster. 

If NGC 1846 is an SSP cluster, our results can exclude the hypothesis that MPs is a specific pattern for low-mass stars \citep{Bast18a}. The lack of MPs in NGC 1846 also supports the conclusions made by 
\cite{Cabr20a} and \cite{Li21a}, based on the analyses of low-mass dwarfs in NGC 419, another intermediate-age Small Magellanic Cloud cluster. MPs is likely a global feature for stars at all evolutionary stages. If a cluster does not exhibit MPs among its evolved stars, it does not have MPs among its non-evolved dwarfs, neither. However, the helium measurements of the present work do not have the accuracy required to detect small helium variations in NGC1846, if present. All the statements above are speculative yet.

We summarize our conclusions as below, 
\begin{itemize}
\item[1.] NGC 1846 is most likely harbor a helium spread of $\Delta{Y}\sim0.01$, or it is a SSP cluster. If NGC 1846 contains no less than 10\% 2G stars, it does not harbor an extreme helium spread ($\Delta{Y}\geq$0.03). 
\item[2.] The estimated $\Delta{Y}$ for NGC 1846 is consistent with the correlation between the maximum helium spread and the clusters total masses derived for Galactic GCs.
\item[3.] Our result cannot determine whether the cluster age is a parameter which controls the presence of MPs, a higher photometric accuracy is required to explore this topic.
\end{itemize}

We also suggest studies focusing on low-mass clusters older than the critical age for the presence of MPs, to determine if cluster mass plays a role in the presence of MPs. It is important to study very massive clusters younger than the critical age, since the role of mass loss of star clusters remains unclear yet. In that case, a deep view of starburst galaxies in the local group will provide helpful insights into the origin of MPs. 

\newpage
\acknowledgements 
{We are grateful for the suggestions from the anonymous referee that have helped improve this manuscript. We thanks Dr. Nicola R. Napolitano at the Sun Yat-sen University for improving the language. C. L. is supported by the National Key R\&D Program of China (2020YFC2201400). C.L. acknowledge support from the one-hundred-talent project of Sun Yat-sen University and the National Natural Science Foundation of China through grant 11803048.}

\end{document}